\documentclass[10pt,prb,twocolumn]{revtex4}
\usepackage{graphicx}

\begin{document}
\title{Morphology of epitaxial core-shell nanowires}

\author{Hailong Wang}
\affiliation{Simulation and Theory of Atomic-scale Material Phenomena, Division of Engineering, Materials Science Program, Colorado School of Mines, Golden, CO 80401}

\author{Moneesh Upmanyu\footnote{Corresponding author, email: mupmanyu@mines.edu}}
\affiliation{Simulation and Theory of Atomic-scale Material Phenomena, Division of Engineering, Materials Science Program, Bioengineering and Life Sciences Program, Colorado School of Mines, Golden, CO 80401}

\author{Cristian V. Ciobanu}
\affiliation{Division of Engineering, Colorado School of Mines, Golden, CO 80401}

\begin{abstract}
We analyze the morphological stability against azimuthal, axial, and general helical perturbations for epitaxial core-shell nanowires in the growth regimes limited by either surface diffusion or evaporation-condensation surface kinetics. For both regimes, we find that geometric parameters (i.e., core radius and shell thickness) play a central role in determining whether the nanowire remains cylindrical or its shell breaks up into epitaxial islands similar to those observed during Stranski-Krastanow growth in thin epilayers. The combination of small cores and rapid growth of the shell emerge as key ingredients for stable  shell growth. Our results provide an explanation for the different core-shell morphologies reported in the Si-Ge system experimentally, and also identify a growth-induced intrinsic mechanism for the formation of helical nanowires.
\end{abstract}

\maketitle
The combination of dimensionally confined transport and engineered interfaces inherent in heterostructured nanowires has led to their emergence as an important class of low dimensional, multifunctional nanostructures.\cite{nw:Lieber:1998,nw:BjorkSamuelson:2002,nw:SamuelsonBjork:2004} In particular, radially heterostructured core-shell nanowires (CSNWs) have enjoyed considerable attention from the synthesis and design communities, mainly due to an unprecedented range of reported electronic properties such as high mobility carrier transport, tunable band gaps, non-linear optical gains, and giant magnetoresistance.\cite{nw:DuanLieber:2001,nw:CuiLieber:2003,nw:DuanAgarwalLieber:2003,nw:ThelanderSamuelson:2003,nw:ChoiJohnson:2003,nw:QianLieber:2004,nw:QianLieber:2005} Modifications to usual nanowire synthesis routes, based on tailoring the rate and chemistry of radial growth during catalysis and/or vapor transport, have allowed the realization of several nanowire systems, including Si/Ge,\cite{nw:LauhonLieber:2002,nw:DickeyRedwing:2005,PNAS2005} GaN/GaP,\cite{nw:LinChen:2003} GaN/InGaN,\cite{nw:QianLieber:2004} GaAs-Ga$_x$In$_{1-x}$P,\cite{nw:SkoldSamuelson:2005} and metal oxides such as SnO$_2$/In$_2$O$_3$.\cite{nw:Sn-ox-In-ox:2007}

\begin{figure}[htp]
\includegraphics[width=\columnwidth]{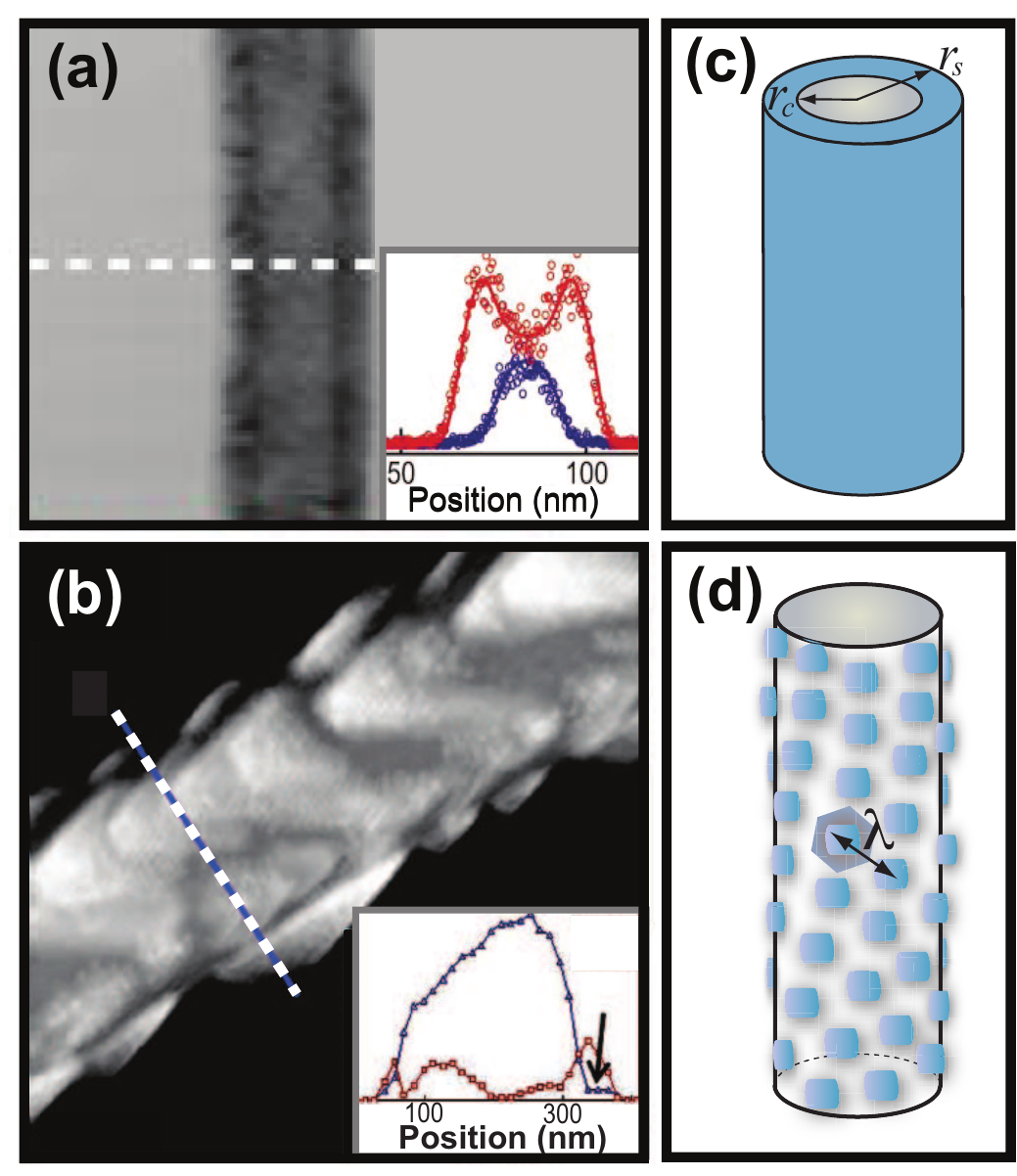}
\caption{Morphologies of epitaxial core/shell nanowires reported in (a) Ref.~\onlinecite{nw:LauhonLieber:2002} and in (b) Ref.~\onlinecite{nw:DickeyRedwing:2005}, for Si-core Ge-shell (Si/Ge) systems with different core radii. We propose that the classical interplay between the mismatch strain at the core-shell interface and surface energy depends on both the core ($r_c$) and shell ($r_s$), such that small fluctuations of the compressively strained Ge shells during  growth can either decay, rendering the nanowire cylindrical (a,c), or can grow and cause the shell to break up into epitaxial islands (b,d). [Permissions to reproduce the viewgraphs in panels (a) and (b) have been requested from the respective authors and publishers, and are currently pending.]}
\label{fig:CSNWExpts}
\end{figure}

The direct control over the size and composition of the core and shell during synthesis should translate to wires with tailored orientations, degree of interface mixing and overall surface structure. Surface morphology is crucial as it drastically modifies the eventual function, yet its control remains an open issue. Indeed, recent studies in epitaxial systems have shown that the wires do not always remain cylindrical but can develop nanoscale surface modulations that tend to self-organize in a manner akin to quantum dots on thin films. For example, consider the Si-Ge system. The experiments of Lauhon {\it et al.} revealed that for small core radii ($r_c\sim10-15$\,nm), the shell grows uniformly in both Si/Ge and Ge/Si nanowires (see Fig.~\ref{fig:CSNWExpts}a).\cite{nw:LauhonLieber:2002} For large core radii ($r_c \geq 50$nm), Pan {\it et al.} found that Ge shells grow non-uniformly  on Si cores and eventually develop into well-defined Ge islands,\cite{nw:DickeyRedwing:2005} as shown in Fig.~\ref{fig:CSNWExpts}b: this observation strongly suggests that the growth occurs in the Stranski-Krastanow mode in which the shell first develops a wetting layer and subsequently forms islands that grow and coarsen. Similar islands morphologies have also been observed in PbSe/PbS\cite{nw:TalapinMurray:2007} and SnO$_2$/In$_2$O$_3$ CSNWs.\cite{nw:Sn-ox-In-ox:2007}

The reports of these intriguing morphologies lead to the question: what conditions determine the stability of the epitaxial shell? In addition to the classical interplay between surface energy and mismatch strain that determines the stability of epilayers,\cite{tsf:atg,tsf:Srolovitz:1989,tsf:VoorheesJohnson:2004}
the shell morphology now also depends on core size and curvature, its growth direction and crystallography. In this letter, we use an isotropic continuum approach to study the combined effect of mismatch strains and geometrical parameters on the morphology of epitaxial CSNWs.
While we concentrate the comparison with experiments mainly on the Si/Ge system, the analysis we carried out is general and provides a
fundamental framework for predicting the influence of geometry on the morphology of any epitaxial core-shell nanowire.

To illustrate the main ideas, consider the scenario where a thin shell of thickness $t=r_s-r_c$ and with average mismatch stress $\bar{\sigma}_s$ breaks up into $n$ stress-free cuboidal islands of volume $v=a\times a\times h$ and average surface energy $\gamma_s$. The energy of nanowire segment of length $l$ is
\begin{eqnarray}
\delta F 
	      &=& - \frac{\bar{\sigma}_s^2}{2E_s} \pi \left(r_s^2 -  r_c^2 \right) l +  n (4 h a\gamma_s),\nonumber
\end{eqnarray}
where the first term is the energy gained via elimination of the elastic strain energy and the second term is the increase in surface area due to island formation. Assuming that the islands self-assemble on the core surface into a hexagonal array with an inter-island distance $\lambda$, as shown schematically in Fig.~\ref{fig:CSNWExpts}d,
the core area occupied by each island is $A\sim
\lambda^2$. The surface area of the core is conserved, i.e. $n A = 2 \pi r_c l$. Setting $\delta F(\lambda)=0$, we find that the preferred inter-island spacing $\lambda_m$ decreases parabolically with shell thickness,
\begin{equation}
\label{eq:crudeAnalysis}
\lambda_m = \frac{\sqrt{8h a \gamma_s E}}{\bar{\sigma}_s} \left( \frac{1}{\sqrt{t}} \right), \quad \quad t \ll r_c.
\end{equation}
Using values for the Si/Ge system ($\gamma_s\approx0.5$\,Jm$^{-2}$, $E\approx1$GPa, $\bar{\sigma}\approx5$\,GPa) and the reported island sizes and spacing in the unstable wires\cite{nw:DickeyRedwing:2005} ($a\approx50$\,nm, $h\approx30$\,nm, $\lambda_m\approx100$\,nm), we extrapolate the maximum thickness of a stable shell to be $t <1$\,nm.

A major oversimplification in the above analysis is that the average stress is assumed to be constant while in reality it also depends on the nanowire geometry (radii $r_s$ and $r_c$) and material parameters (core and shell lattice parameters $a_s$ and $a_c$, Young's moduli $E_s$ and $E_c$). The analysis also does not make any connection with deposition conditions, in particular the surface flux. In the remainder of this article, we present results of a more rigorous linear stability analysis to study the morphological evolution of a dislocation-free, strained shell. Small perturbations in the otherwise cylindrical shell and the associated change in elastic stress distributions are analyzed for stability with respect to surface relaxation kinetics. The elastic solutions are quite detailed and we only summarize the main results in the text; the detailed derivations will be presented elsewhere [HW, MU and CC, in preparation]. For clarity, the results are limited to Si-Ge nanowires although the qualitative trends apply to any epitaxial CSNW system.
\begin{figure*}[htp]
\includegraphics[width=1.4\columnwidth]{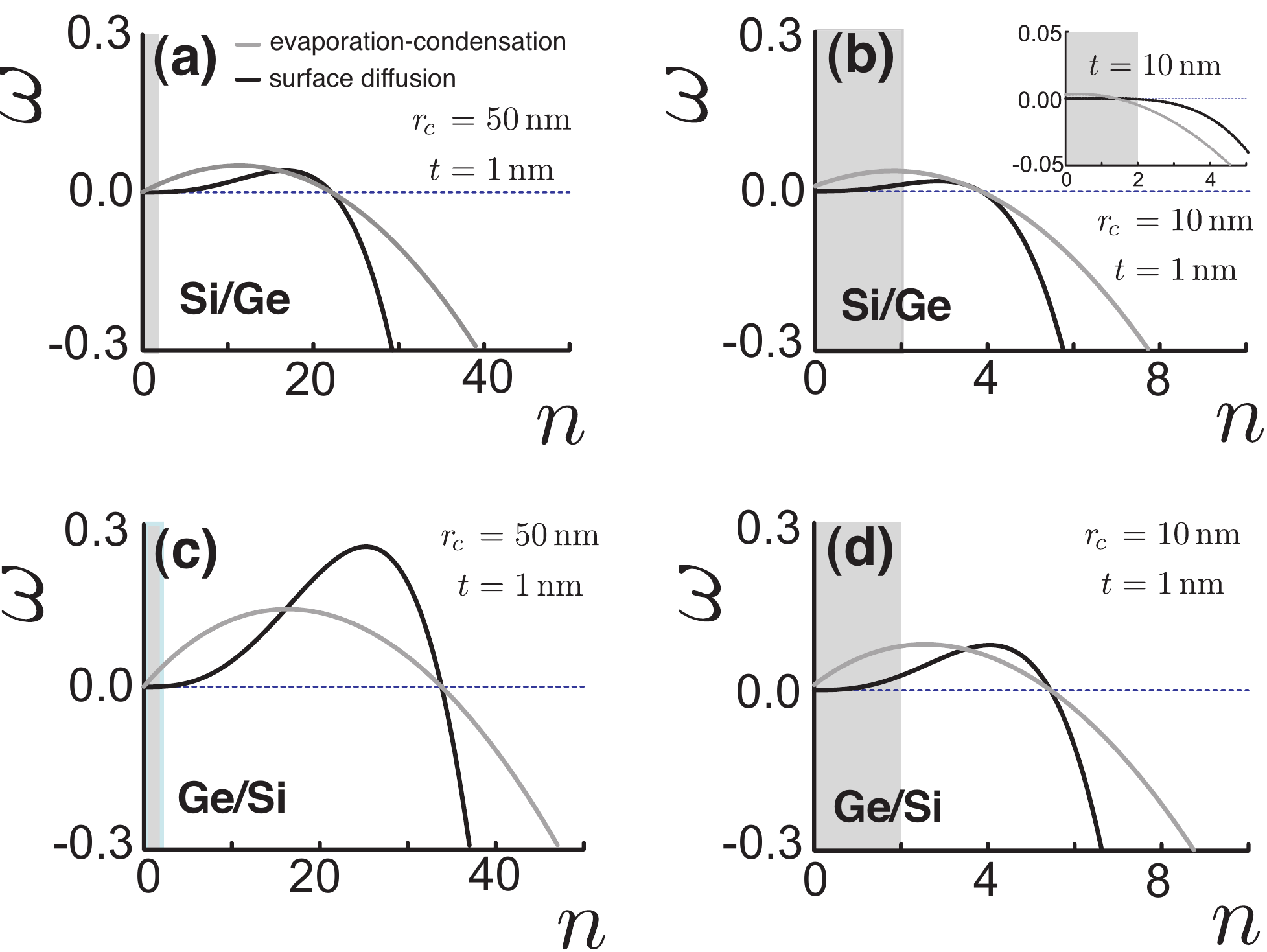}
\caption{(a--d) Dispersion relations $\omega (n)$ for the azimuthal perturbations of the shell when the growth kinetics is limited by surface diffusion (black curves), or by evaporation-condensation (gray curves). The growth rates $\omega$ are in units of $(Bt)^{-1}$. The geometrically forbidden regime of $n<2$ is shown as a vertical gray stripe in each case. The dispersion relations for the Si/Ge systems with $t=1$ nm thick shell are shown in panels (a) and (b) for core radii of $r_c=50$ nm and $r_c = 10$ nm, respectively; panels (c) and (d) show the dispersion relations for Ge/Si systems with the same geometrical parameters as in (a) and (b). The inset in (b) illustrates the stabilizing effect of increasing the shell thickness by a factor of 10.
}
\label{fig:aziDispRelSiGe}
\end{figure*}

Since our focus is on the effect of geometry, we (i) ignore intermixing at the core-shell interface, (ii) approximate the core and shell as isotropic elastic media, and (iii) assume that no facets exist at the core-shell interface. Then, the logical reference for the linear stability analysis is the elastic state of cylindrical CSNWs determined by the dilatational mismatch strain $(a_s - a_c)/a_c$ at the core-shell interface, $r=r_c$. For simplicity, we ignore differences in Poisson's ratio $\nu$ of the core and shell. The strain distribution in the core and shell follow from the displacements  (per unit length), which depend on the geometry and the mismatch. 
Ignoring non-linear elastic effects, Hooke's law and equilibrium conditions readily yield the stress distribution.\cite{nw:OvidkoSheinerman:2004, nw:LiangNix:2005}
Axial symmetry dictates that the variations in stresses and strains are limited to in-plane normal components within the shell. Therefore, the surface stress state is entirely due to the non-zero normal components with similar magnitudes,  $\sigma_{\theta\theta}^\ast$ and $\sigma_{zz}^\ast$. Taken together, they set the energy per atom on the surface, $\mathcal{E}^\ast$. We must emphasize that this surface elastic energy decreases rapidly with the shell thickness as it approaches the dimensions of the core, a trend that we shall later show to have a dominant effect on the morphological stability of the shell.

To make the stability analysis more transparent, we first consider azimuthal perturbations. The nanowire surface geometry now takes the form $r(\theta) = r_s + \Delta \cos n \theta$, where $\theta$ is the (in-plane) angular coordinate, $\Delta$ is the perturbation amplitude, and $n=2\pi/(\lambda/r_s)\ge2$ is the azimuthal wavenumber. Geometrical constraints limit the possible wavenumbers to $n\ge2$. The solution to this plane strain problem is the unperturbed, cylindrical core-shell system superposed with appropriate boundary conditions on an interior shell surface defined by the prescribed perturbation. The stress state is extracted using Airy stress functions for the core and the shell (Mitchell solution)\cite{book:Barber:2002}
and imposing interface conditions at $r=r_c$. The traction free conditions at the surface are absorbed by transforming the stress tensor into local normal ($\eta$)-tangent ($\tau$) frame\cite{tsf:Srolovitz:1989}, and setting $\sigma_{\eta\eta}=\sigma_{\eta\tau}=0$. The equations are solved for small slopes and amplitudes, $n \Delta/r_s\ll 1$.

The resultant stress driven morphological evolution is analyzed at the shell surface with the volume diffusion ignored. The central quantity which drives the relaxation is the surface chemical potential, which we define with respect to that of the cylindrical nanowire, $\mu_s^\ast$. In order to capture the essential physics, we limit ourselves to an elastocapillary potential, $\mu_s=\mu_s^\ast + \Omega \gamma_s \kappa + (\mathcal{E}-\mathcal{E}^\ast)$, where $\Omega$ is the atomic volume in the shell, $\gamma_s$ is the surface energy, $\kappa$ is the surface curvature, and $\mathcal{E}$ is the misfit strain induced elastic energy (per atom) on the surface.\cite{tsf:VoorheesJohnson:2004} While non-linear effects induced by wetting potentials and surface stresses can certainly become important,\cite{tsf:SavinaVoorheesDavis:2004} they are the focus of follow-up work and ignored in this study. We limit ourselves to the two limiting cases of surface kinetics, i.e. surface diffusion and vapor transport via evaporation-condensation at the surface.\cite{gbg:Mullins:1957,tsf:Srolovitz:1989} In the absence of an external deposition flux or at small deposition rates, surface diffusion is expected to control the relaxation (SD-limited kinetics). The flux in this case is proportional to gradient in chemical potential at the surface, $\nabla_s\mu_s$. In the presence of an external surface flux, as in shell growth via chemical vapor deposition (CVD), the morphological evolution is limited by evaporation-condensation at the surface (EC-limited kinetics). At near equilibrium, the flux is proportional to the difference between ambient and local surface vapor pressure, which in turn depends on the local curvature.

In both scenarios, the surface stress state of the perturbed nanowire modifies the surface flux and therefore the radial growth of the shell. In the small slope limit, the governing partial differential equations yield a perturbation amplitude which varies exponentially with time, $\Delta(t) = \Delta(0) \exp[B\omega t]$, where $B$ is a positive constant that depends on the material properties, deposition conditions and surface kinetics.\cite{gbg:Mullins:1957} Clearly, it is the sign of $\omega$ which determines whether the perturbation grows (unstable kinetics) or decays away (stable kinetics). For the azimuthal perturbations, this reduced growth can be expressed as
\begin{equation}
\label{eq:dispRel}
\omega = f\left( \frac{n}{r_s} \right) \left[-\gamma_s \frac{n^2 - 1}{n r_s} + \frac{M(1-\nu^2)}{E_s} (\sigma_{\theta\theta}^*)^2 \right],
\end{equation}
where $M$ is a positive function that depends on $n$, geometry, and material properties. Since the general function is rather complicated involving powers of $n$, in the remainder of this study we investigate its effect numerically for the Si-Ge system. The function $f(n/r_s)\equiv f(\lambda)$ is positive and depends on the type of surface kinetics; it scales as $\lambda^{-3}$ and  $\lambda^{-1}$ for SD- and EC-limited kinetics, respectively. The remaining expression (inside brackets of Eq.~\ref{eq:dispRel}) is the critical condition since its sign determines the stability of the system. The first term captures the stabilization due to high surface energies and surface curvature ($\kappa\propto1/r_s$), while the effect of elastic stresses is embodied in the second term. Setting $\omega=0$, we arrive at the value of the critical perturbation wavenumber $n_{cr}$, which is identical for the two classes of surface kinetics.\cite{tsf:Srolovitz:1989} A direct comparison, however, cannot be made between the two as the growth of the perturbation amplitude $\Delta(t)$ also depends on deposition parameters, lumped in the constant $B$, which are different for the two cases.\cite{gbg:Mullins:1957}

Quantitative insight into the nanowire stability can be obtained from the variation in the reduced growth rate with the perturbation wavenumber, $\omega$ vs $n$, or the dispersion relations. The plots are shown in Fig.~\ref{fig:aziDispRelSiGe} for Si/Ge and Ge/Si systems and for two different core radii,  $r_c=50$\,nm and $r_c=10$\,nm. The choice is based on the nanowire geometries reported in Refs.~\onlinecite{nw:LauhonLieber:2002} and~\onlinecite{nw:DickeyRedwing:2005} (see Fig.~\ref{fig:CSNWExpts}). In each case, the results are reported for a shell of thickness $t=1$\,nm. We find a dramatic effect of the core radius on the range of unstable wavenumbers. In the case of the Si/Ge system, the critical wavenumber for the larger cores is $n_{cr}\approx23$ (Fig.~\ref{fig:aziDispRelSiGe}a). The corresponding azimuthal diffusion length is $\lambda\approx14$\,nm. The prefactor $f(\lambda)$ leads to surface kinetics dependent maximal growth modes $n_{m}$ (value of $n$ for which $\partial \omega/\partial n=0$, also the most unstable mode), quite like the trends reported for stressed thin films.\cite{tsf:Srolovitz:1989,tsf:VoorheesJohnson:2004} For EC-limited kinetics, the maximal wavenumber is $n_m=10$, and the azimuthal (diffusion) wavelength associated with this instability $\lambda_m\approx31$\,nm. This length is smaller for SD-limited kinetics, $\lambda_m\approx17$\,nm.

Figure~\ref{fig:aziDispRelSiGe}b reveals that the range of unstable wavenumbers and the corresponding reduced growth rates $\omega$ are significantly smaller than those corresponding to $r_c=50$\,nm (Fig.~\ref{fig:aziDispRelSiGe}a). While the diffusion lengths do not change, they are now comparable to the core radius, approaching half the circumference of the core cross-section. For example, $n_m\approx3$ during SD-limited kinetics (Fig.~\ref{fig:aziDispRelSiGe}b), resulting in large relative diffusion length,  $\lambda_m/r_s\approx2$. We expect the kinetics to be further retarded as the diffusion lengths are now spread over much larger variation in core curvature such that atomic-scale effects such as barriers at surface steps start to become important.\cite{tsf:EhlrichHudda:1966} The  same trends are predicted for the Ge/Si system (Figs.~\ref{fig:aziDispRelSiGe}c and \ref{fig:aziDispRelSiGe}d), illustrating again that at constant shell thickness $t$  a small value of core radius is a key factor for the 
stability of the cylindrical shell. The parameters considered in Fig.~\ref{fig:aziDispRelSiGe}d ($r_c=10$~nm, $t=1$~nm) are quite similar to those corresponding to the Ge/Si CSNWs recently synthesized by Lu {\em et al.}\cite{PNAS2005}, $r_c=7.5$~nm and $t=$2--5~nm. Using the experimental values, we have found that the diffusion wavelength is about two times larger than the core radius. Our analysis thus predicts that the Si shell of Ge/Si CSNWs is stable, which is consistent with the experiments of Lu {\em et al.}\cite{PNAS2005} The effect of a stiffer Si shell under tensile strain is to increase the range of unstable wavenumbers and the corresponding growth rates. Again, however, strain induced thermodynamic and kinetic effects are not factored in our analysis and they could become important for stabilizing the tensile Si epilayers in these nanowires.\cite{tsf:XieGilmerCitrin:1994}

A similar approach is employed to investigate shell stability with respect to axial perturbations. The shell surface morphology depends on the axial coordinate, $z$. We have considered perturbations of the form $r(z)= r_s + \Delta \cos kz$, where $k$ is the axial wavenumber. The elastic solution can be expressed in terms of Papkovich-Neuber potentials.\cite{book:Barber:2002} Since the core and shell have different elastic moduli, the relevant Airy stress function now involves modified Bessel functions of the first and second kind. The linear stability analysis reveals that the dispersion relations, $\omega$ vs $kr_s$, are similar qualitatively to those corresponding to azimuthal perturbations. This is not surprising and merely indicates that trends dictated by the balance between surface and elastic energies, calculated in the isotropic limit, are approximately the same for axial and azimuthal perturbations; the effect of inherent anisotropy in geometry on stability of thin shells that we have considered is relatively small.

\begin{figure}[htp]
\includegraphics[width=\columnwidth]{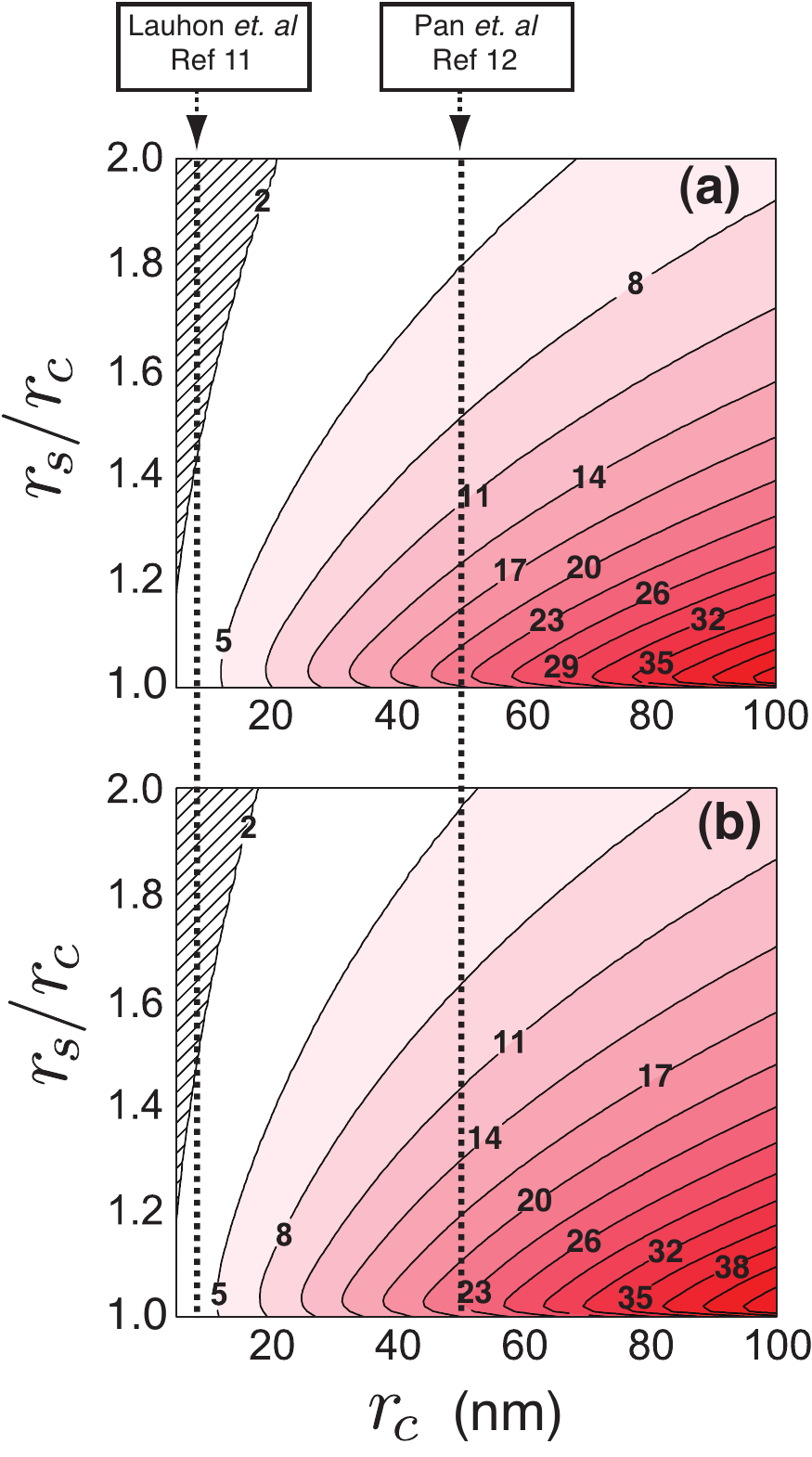}
\caption{Contour plots of (a) the critical azimuthal wavenumbers $n_{cr}$ and (b) the critical axial wavenumbers $k_{cr}r_s$ associated with surface diffusion-limited  relaxation of a sinusoidally perturbed Ge shell in a Si/Ge nanowire, as functions of the core and shell radii, $r_s$ and $r_c$ respectively. The hashed region indicates a stable shell. Note that the definition of a stability with respect to axial perturbations is somewhat arbitrary as the perturbation is limited by the length of the wire. The cut-off wavenumber below which we consider the nanowire to be stable corresponds to a nanowire length of approximately $1\,\mu$m. The vertical dotted lines correspond to growth on Si cores with radii $r_c$ reported in Refs.~\onlinecite{nw:LauhonLieber:2002} and \onlinecite{nw:DickeyRedwing:2005}, as indicated.}
\label{fig:aziModulusEffect}
\end{figure}

The inset in Fig.~\ref{fig:aziDispRelSiGe}b shows the dispersion relation for azimuthal perturbations in the Si/Ge nanowire for a ten-fold increase in shell thickness ($t=10$\,nm). The shell has the same dimensions as the core, $t=r_s$. Clearly, the increase in shell thickness has a remarkable effect on the stability. The growth rate is now negative for all possible perturbations indicating that at least for the small core sizes employed in the experiments of Lauhon {\it et al.}, there exists a critical shell thickness beyond which the shell becomes stable. In order to ascertain if this is a general trend, we have systematically analyzed the stability of Si/Ge nanowires with various core and shell sizes. The results are presented in the form of stability diagrams -- contour plots of the critical azimuthal ($n_{cr}$) and axial wavenumbers ($k_{cr}r_s$) as functions of nanowire geometry (Fig.~\ref{fig:aziModulusEffect}). For large core radii, the sharp turning point (knee) in the contours indicates that thin shells ($r_s/r_c$ ratio below the knee) become progressively unstable initially. Interestingly, further growth past the knee reverses this trend. The enhanced stability is more pronounced for small core radii, where the contours no longer exhibit a turning point. Increasing the shell thickness always increases its stability with respect to both azimuthal and axial perturbations. 

It is informative to discuss the predictions of stability phase diagrams such as those in Fig.~\ref{fig:aziModulusEffect} for other systems than Si/Ge CSNWs. We have found that for the SnO$_2$/In$_2$O$_3$ (Ref.~\onlinecite{nw:Sn-ox-In-ox:2007}) and GaN/GaP (Ref.~\onlinecite{nw:LinChen:2003}) nanowire systems the experimental observations are in close agreement with our theoretical stability analysis. The phase diagrams for these two systems are similar to those given in Fig.~\ref{fig:aziModulusEffect} for Si/Ge nanowires. The GaN/GaP system has a uniform cylindrical stable shell: this is consistent with our theory, which based on the geometric parameters of Ref.~\onlinecite{nw:LinChen:2003} ($r_c=7$~nm, $r_s/r_c = 3.0$) predicts that the GaN/GaP system is deep inside the stable region for both azimuthal and axial perturbations. For the SnO$_2$/In$_2$O$_3$ system,\cite{nw:Sn-ox-In-ox:2007} there are two distinct morphological regimes observed experimentally, both of which agree with the theoretical analysis. The SnO$_2$/In$_2$O$_3$ CSNWs with smaller core radii and larger $r_s/r_c$ ratios (e.g., $r_c=33$~nm, $r_s/r_c=1.76$) are cylindrical and have a stable shell,\cite{nw:Sn-ox-In-ox:2007} while the shell of the nanowires with larger core radii and smaller $r_s/r_c$ ratios (e.g. $r_c=45$~nm, $r_s/r_c=1.64$) is unstable\cite{nw:Sn-ox-In-ox:2007} and develops islands similar to those reported by Pan {\em et al.}\cite{nw:DickeyRedwing:2005} Lastly, we note that for Ge-core/Si-shell systems, the softer Ge cores result in Si shells which increase their stability irrespective of the core sizes considered ($<100$\,nm), i.e. there is no knee in the contours. The results of the stability analysis are again consistent with the observations of stable shell Ge/Si CSNWs by Lu {\em et al.}\cite{PNAS2005}, since the experimental parameters place these Ge/Si nanowires close to the stability boundary (not shown) in our calculations.

The trend is opposite to that predicted by Eq.~\ref{eq:crudeAnalysis} ($\lambda \propto 1/\sqrt{t}$), and also to that normally observed in compressively strained thin films on planar substrates.\cite{tsf:Srolovitz:1989,tsf:VoorheesJohnson:2004} In these core-shell heterostructures with finite-size core, the elastic stresses and surface (shell) curvature both decrease rapidly with increasing shell thickness.
The stability diagrams show that the former is more sensitive as the shell approaches the dimensions of the core. The destabilization due to decrease in surface curvature ($\sim\gamma_s/r_s$, the first term in the parenthesis in Eq.~\ref{eq:dispRel}) is relatively small, and overall the shell becomes more stable. The behavior underscores the importance of a finite core in absorbing the mismatch stresses in these low dimensional heterostructures, thereby dramatically changing the balance between misfit and surface stresses that is at the heart of the Asaro-Tiller-Grinfeld instability in thin films.\cite{tsf:atg}

\begin{figure*}[h!tp]
\includegraphics[width=2\columnwidth]{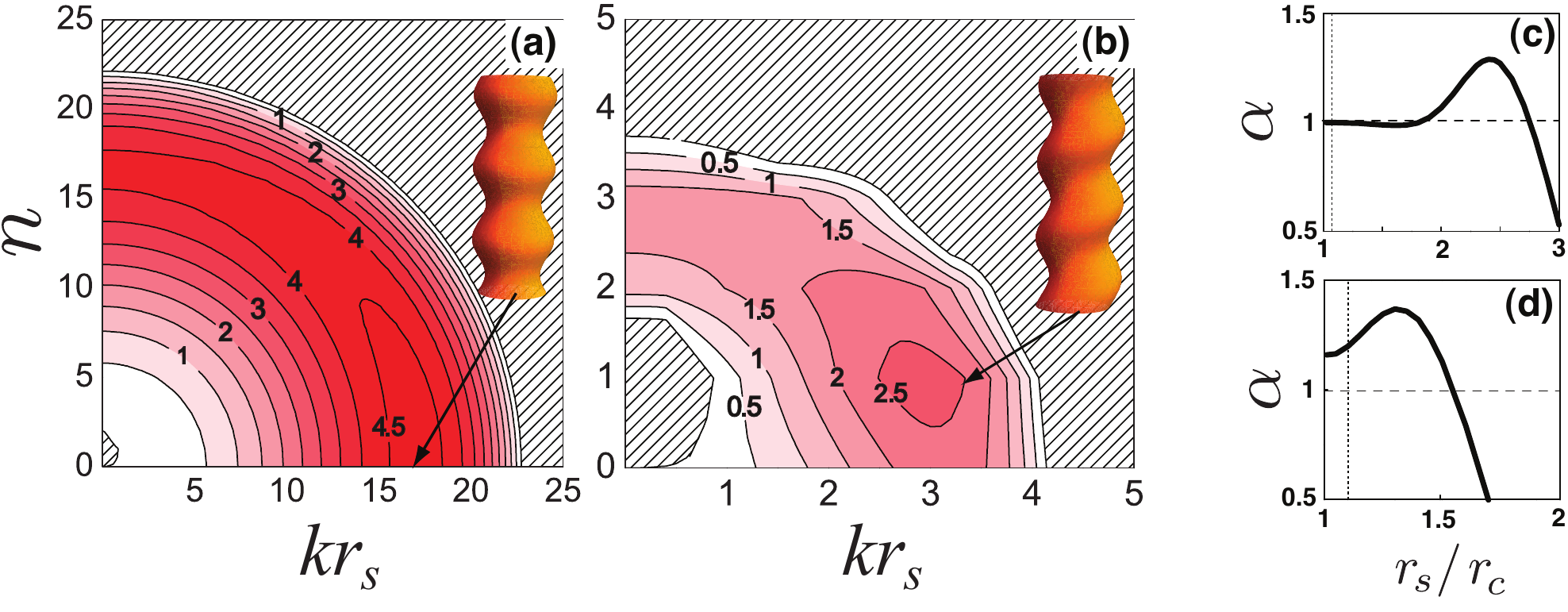}
\caption{Contour plots of the reduced growth rate $\omega(n, kr_s)$ a function of the azimuthal and axial wavenumbers for Si/Ge nanowires with two different geometries: (a) $r_c=50$\,nm, $r_s=51$\,nm, and (b) $r_c=10$\,nm, $r_s=11$\,nm. The growth rates are in given units of $10^{-2}\,(Bt)^{-1}$. The plots in (c) and (d) show the ratio $\alpha$ of the maximally growing helical to axial growth rates, $\alpha = \omega_m (1, k r_s)/\omega_m (0,kr_s)$, for the $k$ values corresponding to the points shown by the arrows in (a) and (b), respectively. The vertical dash lines in (c) and (d)
correspond to the $r_s/r_c$ values that were used in panels (a) and (b), respectively.}
\label{fig:aziAxialGrowthRate}
\end{figure*}

The stability diagrams provide an elegant yet simple explanation for the variations in the morphologies of the as-synthesized CSNWs. To directly make contact with the Si-Ge system, we have overlaid the approximate growth path of the Ge shell in the two studies shown in Fig.~\ref{fig:CSNWExpts}. For the small Si core employed by Lauhon {\it et al.},\cite{nw:LauhonLieber:2002}
the window over which the compressively strained Ge shell is morphologically unstable is small (below the hashed region) and limited to early stages of growth, which explains the uniform, cylindrical shell morphology observed in these CSNWs. In addition to the balance between surface curvature and the elastic stresses, both of which decay with shell growth (Eq.~\ref{eq:dispRel}), we expect stabilization due to the kinetic barriers to the surface relaxation that now needs to occur over larger lengths and surface curvatures. On the other hand, CSNWs with thicker Si cores reported by Pan {\it et al.}\cite{nw:DickeyRedwing:2005} result in a larger window of shell thicknesses over which it can develop axial or azimuthal instabilities. The range of unstable wavenumbers during early stage growth correspond to smaller diffusion lengths. The morphological instability is further aided by the reduced changes in curvature spanned by the smaller diffusion lengths, and also by the higher temperatures ($\sim770\,^\circ$K) employed for the shell growth. The latter follows from the fact that the uncatalyzed CVD-based synthesis route for Ge shell deposition exploits the lower thermal stability of the precusor gas (GeH$_4$), as opposed to the lower downstream temperatures ($\sim700\,^\circ$K) prevalent during stable shell growth in the study by Lauhon {\it et al.}. The combination of these factors makes the shell susceptible to island growth during early stages of its growth, which is precisely what is observed in the experiments.\cite{nw:DickeyRedwing:2005}

Closer examination of the reported island morphologies in the unstable Si/Ge nanowires reveals an axial inter-island distance in the range $\lambda_m=75-150$\,nm. The axial diffusion wavelength length associated with the most unstable mode predicted by our analysis is $\lambda_m=31$\,nm (EC-limited kinetics). The quantitative disparity is hardly surprising as we ignore effects such as intermixing and formation of misfit dislocations at the core-shell interface, growth crystallography, surface stresses, and faceting of the core. Faceting in particular appears to be quite important, as the islands are observed to form preferentially on planar facets; the stress relaxation at the corners results in enhanced stabilization, consistent with Eq.~\ref{eq:dispRel}.

We have also analyzed the interplay between the two perturbation modes, by considering general perturbations of the form $r(\theta, z)= r_s + \Delta \cos(n\theta + kz)$. Figure~\ref{fig:aziAxialGrowthRate} shows combined effect of the two modes on SD-limited reduced growth rate $\omega$ in the Si/Ge system, again for a shell of thickness $t=1$\,nm. The semi-circular contours  follow from the fact that the effective wavenumber is a linear combination of the two modes, $k_{eff}=\sqrt{k^2r_s^2+n^2}$. For large cores (Fig.~\ref{fig:aziAxialGrowthRate}a), the most unstable mode is clearly along the axial direction. Superposition of an azimuthal component always decreases $\omega$, indicative of an intrinsic stabilization mechanism that is likely to be activated due to the spatial variations that naturally exist during the uncatalyzed deposition of the shell. The extent of decrease depends on the magnitude of the scaling factor for the growth rate, $B$. While the islands do form axially in the corresponding experiments,\cite{nw:DickeyRedwing:2005} they cannot serve as validation since the Si core is distinctly faceted and therefore influences the interplay between the two perturbation modes.

For smaller cores, the reduced growth rate exhibits a maximum for the general perturbation $n=1, kr_s\approx3$ (Fig.~\ref{fig:aziAxialGrowthRate}b). The helical radial growth of the shell, shown schematically in the figure, is simply a reflection of the energy balance captured in Eq.~\ref{eq:dispRel}. The capillary and elastic energy terms are functions of the effective wavenumber such that, per unit length of the nanowire, addition of a helical component to an otherwise axial perturbation decreases the curvature dependent surface energy and/or increases the stored elastic energy in the shell. The geometrical window associated with these unstable helical perturbations changes significantly with the growth of the shell, primarily due to the variation of the elastic energy (term $M$ in Eq.~\ref{eq:dispRel}) with the effective wavenumber $k_{eff}$. As an illustration, Figs.~\ref{fig:aziAxialGrowthRate}c and \ref{fig:aziAxialGrowthRate}d show the effect of an azimuthal component ($n=1$) on the maximally growing axial perturbation $\omega_m$, as a function of the geometric ratio $r_s/r_c$. For larger cores (Fig.~\ref{fig:aziAxialGrowthRate}c), the ratio of the reduced growth rates $\alpha =  \omega_m (1, k r_s)/\omega_m (0,kr_s)$ during initial stages of shell growth is slightly less then unity, consistent with the stability diagram reported for $t=1$\,nm. As the shell thickness increases, we observe a window of radii ratio for which helical perturbations should dominate the growth morphology, $2\le r_s/r_c \le 2.7$. In the case of the smaller cores, the ratio is greater than unity during early stages. However, the helical growth mode switches to a pure axial instability past a critical radii ratio, $r_s/r_c\approx1.6$: this variation underscores the importance of geometry on the evolution of the shell morphology.

Krill {\it et al.} have predicted the stability of such helical morphologies via surface diffusion, albeit for pure whiskers subject to extrinsic strains.\cite{nw:Northwestern:recent} In the case of epitaxial core-shell architecture where the mismatch strains are present intrinsically, it may become prohibitive for the core to sustain the helical growth of the shell elastically. A likely scenario, especially when the shell is softer than the core, is that the core buckles helically to better accommodate the perturbations of the shell. The net twist entailed by the buckled core can also be relaxed by defects such as screw dislocations. The fact that such morphologies can grow due to mismatch strains (and possibly surface stresses) appears to be a novel intrinsic mechanism for stability of helical morphologies in core-shell nanowire systems.

In summary, the results of the linear stability analyses for epitaxial core-shell nanowire systems highlight the importance of geometry in determining the nanowire morphology. The dependence on the core size allows us to rationalize the cylindrical (stable) and island (unstable) growth observed experimentally in the Si-Ge system. The variation with shell size shows the importance of the growth rate, relative to surface relaxation kinetics, in controlling the shell morphology. Since the temperature and partial pressure of the precursor gas directly affect the growth rate, they become important deposition variables that allow control over the nanowire morphology. There are, for example, CSNWs systems\cite{nw:TalapinMurray:2007} 
whose morphology can be changed from stable shells to epitaxial islands solely by adjusting the growth rate, even in regimes of geometric 
parameters where our theory strictly predicts uniform shells. In future studies, we will envision the development of a more comprehensive understanding of the morphological evolution of core-shell nanowire heterostructures in which the effects of intermixing, anisotropic surface energies, surface stresses, and kinetics could be addressed in various combinations.

{\bf Acknowledgments.} MU acknowledges support from U.S. Department of Energy (DOE)-sponsored Computational Materials Science Network (CMSN) on ``{\it Dynamics and Cohesion of Materials Interfaces and Confined Phases Under Stress}". MU and HL also acknowledge support from the Nanoscale Science and Engineering Center (NSEC) on ``{\it High-Rate Nanomanufacturing}" at Northeastern University, where parts of this work were carried out.

\end{document}